\documentstyle{article}
\newcommand{\be}{\begin{equation}}
\newcommand{\ee}{\end{equation}}
\newcommand{\bea}{\begin{eqnarray}}
\newcommand{\eea}{\end{eqnarray}}
\newcommand{\ba}{\begin{array}}
\newcommand{\ea}{\end{array}}

\def\bbox{{\,\lower0.9pt\vbox{\hrule \hbox{\vrule height 0.2 cm
\hskip 0.2 cm \vrule height 0.2 cm}\hrule}\,}}
\newcommand{\dsl}{\pa \kern-0.5em /}

\newcommand{\nn}{\nonumber \\}
\newcommand{\tr}{{\rm tr}\,}

\def\ben{\begin{equation}}
\def\een{\end{equation}}
\def\bena{\begin{eqnarray}}
\def\eena{\end{eqnarray}}

\def\today{\ifcase\month\or
  January\or February\or March\or April\or May\or June\or
  July\or August\or September\or October\or November\or December\fi
 \space\number\day, \number\year}

\input amssym.def
\input amssym.tex
\font\mybb=msbm10 at 10pt
\def\bb#1{\hbox{\mybb#1}}
\def\bZ {\bb{Z}}
\def\bR {\bb{R}}
\def\bE {\bb{E}}

\def\bH {\bb{H}}
\def\bC {\bb{C}}

\def\bO {\bb{O}}

\begin{document}


\begin{titlepage}
\vfill
\begin{flushright}
QMW-PH-99-14\\
DAMTP-1999-137\\
LPTENS 99153\\
hep-th/0001024\\
\end{flushright}


\vfill

\begin{center}
\baselineskip=16pt
{\Large\bf BPS states of D=4 N=1 supersymmetry}
\vskip 0.3cm
{\large {\sl }}
\vskip 10.mm
{\bf ~Jerome P. Gauntlett$^{*,1}$, ~{}Gary W. Gibbons$^{\dagger,\flat,+,2}$,
{}~Christopher M. Hull$^{*,\sharp,3}$ and  ~Paul K. Townsend$^{+,4}$ } \\
\vskip 1cm
{\small
$^*$
  Department of Physics\\
  Queen Mary and Westfield College,\\
  Mile End Rd, London E1 4NS, UK\\
}
\vspace{6pt}
{\small
$^\dagger$
  Laboratoire de Physique Th{\'e}orique,\\
Ecole Normale Sup{\'e}rieure,\\
24 Rue Lhomond, Paris 05, France\\
}
\vspace{6pt}
{\small
$^\flat$
Yukawa Institute for Theoretical Physics,\\
Kyoto University,\\
Kyoto 606-8502, Japan\\
}
\vspace{6pt}
{\small
$^\sharp$
Institute for Theoretical Physics\\
University of California Santa Barbara\\
CA 93106-4030, USA\\
}
\vspace{6pt}
{\small
 $^+$
DAMTP,\\
Centre for Mathematical Sciences, \\
Wilberforce Road, \\
Cambridge CB3 0WA, UK\\
}
\end{center}
\vfill
\par
\begin{center}
{\bf ABSTRACT}
\end{center}
\begin{quote}

We find the combinations of momentum and domain-wall charges corresponding to
BPS states preserving  1/4, 1/2 or 3/4 of D=4 N=1 supersymmetry,
and we show how the supersymmetry algebra implies their stability. These
states form the boundary of the convex cone associated with 
the Jordan algebra of $4\times 4$ real symmetric matrices, 
and we explore some implications of
the associated geometry. For the Wess-Zumino model we derive the conditions
for preservation of 1/4 supersymmetry when one of two parallel domain-walls
is rotated and in addition show that this model does not admit any classical
configurations with 3/4 supersymmetry. Our analysis also provides information
about BPS states of N=1 D=4  anti-de Sitter supersymmetry.

\vfill
 \hrule width 5.cm
\vskip 2.mm
{\small
\noindent $^1$ E-mail: j.p.gauntlett@qmw.ac.uk \\
\noindent $^2$ E-mail: g.w.gibbons@damtp.cam.ac.uk \\
\noindent $^3$ E-mail: c.m.hull@qmw.ac.uk\\
\noindent $^4$ E-mail: p.k.townsend@damtp.cam.ac.uk \\
}
\end{quote}
\end{titlepage}
\setcounter{equation}{0}
\section{Introduction}

Although   N=1 supersymmetric field theories in $3+1 $ dimensions have
been extensively investigated for more than twenty five years, most of these
investigations have been based on the standard supersymmetry algebra. It has
been known for some time, however, that $p$-brane solitons in supersymmetric
theories carry $p$-form charges that appear as central charges in the spacetime
supertranslation algebra \cite{az}. Allowing for all such charges,
the D=4 N=1 supertranslation algebra is spanned by a four component Majorana
spinor charge $Q$, the 4-vector $P_\mu$ and a Lorentz 2-form charge
$Z_{\mu\nu}$. The only non-trivial relation is the anticommutator
\be\label{susyalgcov}
\{Q,Q\} = C \gamma ^\mu P_\mu + { 1\over 2} C\gamma^{\mu \nu}Z_{\mu \nu},
\ee
where $C$ is the charge conjugation matrix and $\gamma_\mu
=(\gamma_0,\gamma_i)$
are the four Dirac matrices. Our metric convention is `mostly plus' so that we
may choose a real representation of the Dirac matrices. In this representation
the Majorana spinor charges $Q$ are real, so $\{Q,Q\}$ is a symmetric $4\times
4$ matrix with a total of ten real entries. The number of components of $P_\mu$
and $Z_{\mu\nu}$ is also ten, so that we have indeed included all possible
bosonic central charges. Note that the automorphism group of this algebra is
$GL(4;\bR)$.

The components of $Z_{\mu\nu}$ can be interpreted as charges carried by
domain walls \cite{az}, while $P_\mu$ is (in general) a linear combination of 
the momentum and a string charge. In the case of a domain wall, the tension is
bounded by the charge, and saturation of this bound implies preservation of 1/2
of the N=1 D=4 supersymmetry. This is one example in the class of
`1/2 supersymmetric' configurations allowed by the supersymmetry
algebra\footnote{An analysis of 1/2 supersymmetric combinations of charges in
$N>1$ D=4 theories, $N=2$ in particular, can be found in \cite{ferrara}.} 
Such 1/2 supersymmetric domain walls were shown to
occur in \cite{AT} in the Wess-Zumino (WZ) model, for an appropriate
superpotential, and also arise in the $SU(n)$ SQCD \cite{DS} because the
low-energy effective Lagrangian is that of a WZ model with a superpotential
admitting $n$ discrete vacua \cite{VY}. More recently, it was shown
that the WZ model also admits (again for an appropriate superpotential) 1/4
supersymmetric configurations that can be interpreted as intersecting domain
walls \cite{GT,CHT}. More precisely, it was established that such configurations
must solve a certain `Bogomol'nyi' equation for which earlier
mathematical studies had made the existence of appropriate solutions plausible
(especially in view of the results of \cite{gui} which were recently brought to
our attention). Domain wall junctions of the WZ model have
since been studied further in \cite{saffin,gorsky,binosi,SV} and an explicit 1/4
supersymmetric domain wall junction of a related model has recently been found
\cite{oda}.

It was pointed out in \cite{GT} that the possibility of 1/4 supersymmetric
intersecting domain walls is inherent in the supersymmetry algebra. If we
choose $C=\gamma^0$ and $\gamma_5=\gamma^0\gamma^1\gamma^2\gamma^3$,
then (\ref{susyalgcov}) becomes
\be\label{susyalg}
\{Q,Q\} = H + \gamma^{0i}P_i + {1\over2}\gamma^{0ij}U_{ij} +
{1\over2}\gamma^{0ij}\gamma_5 V_{ij}
\ee
where $H=P^0$, $U_{ij}= Z_{ij}$ and $V_{ij}= -\varepsilon_{ijk}Z_{0k}$.
One is thus led to expect `electric' type domain walls with non-zero
2-form $U_{ij}$ but vanishing $V_{ij}$ and `magnetic' type domain walls with
non-zero 2-form $V_{ij}$ but vanishing $U_{ij}$. In general, a domain wall will
be specified not only by its tension and orientation but also by an angle in
the electric-magnetic charge space; the domain wall is `dyonic' when this angle
is not a multiple of $\pi$. It is not difficult to show that the algebra
(\ref{susyalg}) allows for dyonic charge configurations preserving 1/4
supersymmetry. In this paper we determine the model-independent restrictions on
such configurations that are implied by the supersymmetry algebra.

As pointed out in \cite{GT}, the charge 
associated with the stringlike junctions of
domain walls in the WZ model appears in the supersymmetry algebra in the same
way as the 3-momentum, so for a static 1/4 supersymmetric configuration of the
WZ model the 3-vector ${\bf P}$ must be interpreted as a string
charge carried by the domain wall junction. It was further shown in \cite{GT}
that this junction charge
contributes positively to the energy of the 1/4 supersymmetric configuration as
a whole. In contrast, the charge associated to domain wall junctions of the
model considered in \cite{oda} was shown there to contribute {\sl negatively}
to the total energy. As we shall see, this apparent disagreement is due to a
different central charge structure for the two models. There is therefore more
than one field theory realization of static intersecting domain walls
preserving 1/4 supersymmetry, but as yet no example that exploits the most
obvious possibility in which ${\bf P}$ vanishes but $U_{ij}$ and $V_{ij}$ do
not.

These observations underscore the importance of the model-independent
analysis of 1/4 supersymmetric configurations based only on the N=1 D=4
supersymmetry algebra, but our aim is to understand the implications of the
supersymmetry algebra for {\sl all} supersymmetric configurations, not just
those preserving 1/4 supersymmetry. Since the matrix $\{Q,Q\}$ is a positive
definite real symmetric one, it can be brought to diagonal form with real
non-negative eigenvalues. The number of zero eigenvalues is the number of
supersymmetries preserved by the configuration. The `supersymmetric'
configurations are those for which this number is $1,2,3$  or $4$. There is a
unique `vacuum' charge configuration preserving all four supersymmetries.
Configurations preserving two supersymmetries are 1/2 supersymmetric while
those preserving one supersymmetry are 1/4 supersymmetric. Configurations
preserving three supersymmetries are 3/4 supersymmetric, but  there is no
known field theoretic realization of this possibility. Indeed, we will show
here that there is no classical field configuration of the WZ model that
preserves 3/4 supersymmetry. However, possible string-theoretic realizations
of exotic supersymmetry fractions such as 3/4 supersymmetry were recently
explored in \cite{GH}, and this possibility has been considered previously in
a variety of other contexts \cite{pope,Bandos,BLS,Ueno}. In particular, the
$OSp(1|8;\bR)$-invariant superparticle model of \cite{Bandos} provides a
simple realization in the context of particle mechanics. The fundamental
representation of $OSp(1|8;\bR)$ is spanned by
$(\rho^\alpha,\lambda_\alpha,\zeta)$, where $\rho$ and $\lambda$ are two
4-component real commuting spinors of $Spin(1,3)$, and $\zeta$ is a real
anticommuting scalar. The action   \be
\label{spac}
S= \int dt \left[ \rho^\alpha \dot \lambda_\alpha +
\zeta\dot\zeta
\right]
\ee
is manifestly $OSp(1|8)$ invariant; in particular, it is supersymmetric with
supersymmetry charge $Q=\lambda\zeta$. The canonical (anti)commutation
relations
imply that $\{Q_\alpha,Q_\beta\}= \lambda_\alpha\lambda_\beta$, which is a
matrix
of rank one, corresponding to 3/4 supersymmetry.

Thus, there exist models of one kind or another in which all possible fractions
of D=4 N=1 supersymmetry are preserved. This fact provides further motivation
for the general model-independent analysis of the possibilities allowed by the
supersymmetry algebra that we present here. As we shall explain, the space of
supersymmetric charge configurations, or `BPS states', is the
boundary of the convex cone of $4\times 4$ real symmetric matrices
and this has an interpretation in terms of  Jordan algebras.
In analogy with the way that the conformal group acts on massless states on the
light-cone  $P^2=0$, there is a group $Sp(8,\bR )$ that acts on the `BPS cone'
of supersymmetric configurations and which  has an interpretation in this
context as the M\"obius group of the Jordan algebra \cite{gunaydin}.
Another purpose of this paper is to explore some of the geometrical ideas
underlying this interpretation of supersymmetric charge configurations.

It is generally appreciated that BPS states are stable states, this being the 
main reason for their importance, but some ``standard'' arguments 
for stability rely on physical intuition derived from 
special cases. For example,
a massive charged particle that minimises the energy for given charge
cannot radiate its energy away in the form of uncharged photons because 
this would leave behind a particle with the same charge but lower energy,
contradicting the statement that the original particle minimised the energy in 
its charge sector. However, this heuristic argument is not conclusive. For 
instance, the stability against radiative relaxation to a lower energy state
of the same `charge vector' assumes that the radiated energy carries away no 
momentum because momentum is one of the charges, and this assumption would be 
violated by a decay in which just one photon is emitted. It is also implicit in
the heuristic argument that prior to decay one can go to the rest frame, but 
the supersymmetry algebra allows BPS states for which this is not possible,
a massless particle being an obvious, but by no means the only, example.
These considerations show that it is not quite as obvious as generally 
supposed that BPS states are stable. Here we provide a complete analysis, for
the general D=4 N=1 supersymmetry algebra, based on a combination of 
the Minkowski reverse-triangle inequality for positive-definite matrices
and the ordinary triangle inequality for BPS energies. 

The supertranslation algebra for which (\ref{susyalgcov}) is the only
non-trivial (anti)commutator is a contraction of the superalgebra
$osp(1|4;\bR)$, which is the D=4 N=1 anti-de Sitter (adS) superalgebra.
The anticommutator of the 4 real supercharges of the latter is
\be\label{ads}
\{Q,Q\} = C \gamma ^\mu P_\mu + { 1\over 2} C\gamma^{\mu \nu}M_{\mu \nu},
\ee
where $M_{\mu\nu}$ are the Lorentz generators. This is formally equivalent
to (\ref{susyalgcov}), although the charges on the right hand side are no
longer central because they generate the adS group $SO(3,2)$.
However the positivity conditions on these charges are the same, as are the
conditions for preservation of supersymmetry. This fact means that much of our
analysis of the centrally-extended supertranslation algebra can be immediately
applied to the adS case. A related analysis has been considered previously
for D=5 in \cite{FP}, where the D=4 case was briefly mentioned, and
BPS states in D=4 adS have also been analysed by other methods in
\cite{Bandos2}. 

We begin with an analysis of the N=1 D=4 supersymmetry algebra, determining
the charge configurations that preserve the various possible fractions of
supersymmetry, and we show how the positivity of $\{Q,Q\}$ implies the
stability of BPS states carrying these charges. We also show how the
supersymmetry algebra determines, in a model-independent way, some properties
of the 1/4 supersymmetric intersecting domain walls that are realized by the WZ
model, but show also that 3/4 supersymmetry is not realized by classical WZ
field configurations. We then turn to an exposition of the geometry associated
with the supersymmetric configurations, which is that of self-dual homogeneous
convex cones, and review their relation to Jordan algebras. We 
then discuss how our results apply to D=4 N=1 adS supersymmetry, and 
conclude with comments on implications and generalizations of our work, in
particular to M-theory. 

\section{BPS states}

The anticommutator (\ref{susyalg}) can be rewritten as
\be\label{susyalgtwo}
\{Q,Q\} = H + \gamma^{0i}P_i + \gamma_5\gamma^iU_i + \gamma^iV_i
\ee
where
\be
U_i= {1\over2}\varepsilon_{ijk}U_{jk}  \, \qquad
V_i = {1\over2}\varepsilon_{ijk} V_{jk}\, .
\ee
As mentioned above, a charge configuration is supersymmetric if the matrix
$\{Q,Q\}$ has at least one zero eigenvalue. Thus, supersymmetric charge
configurations are those for which $\{Q,Q\}$ has vanishing determinant. We see
from (\ref{susyalgtwo}) that this determinant must be expressible in terms of
$H$ and the three 3-vectors ${\bf P}$, ${\bf U}$ and ${\bf V}$.

Now $\det\{Q,Q\}$ is manifestly $SL(4;\bR)$ invariant, but the subgroup that
keeps $H$ fixed is its maximal compact $SO(4)\cong [SU(2)\times
SU(2)_R]/\bZ_2$ subgroup. Ignoring the quotient by $\bZ_2$, the
first $SU(2)$ factor can be identified with the 3-space rotation group
while the $SU(2)_R$
group rotates the three 3-vectors ${\bf P}$, ${\bf U}$ and ${\bf V}$ into
each other, i.e. these three 3-vectors form a triplet of $SU(2)_R$. The
notation chosen here reflects the fact that $SU(2)_R\supset U(1)_R$,
where $U(1)_R$ is the R-symmetry
group\footnote{This symmetry is usually broken in D=4 N=1 QFTs, either by the
superpotential or by anomalies. We shall comment on this fact in 
the conclusions, but it is not relevant to the purely algebraic analysis 
presented here.} rotating
${\bf U}$ into ${\bf V}$ keeping ${\bf P}$ fixed
(this is the automorphism group of the
standard supersymmetry algebra, including Lorentz generators).
We conclude from this that $\det\{Q,Q\}$ is a
fourth-order polynomial in $H$ with coefficients that are homogeneous
polynomials in the three algebraically-independent $SU(2)\times
SU(2)_R$ invariants that can be constructed from ${\bf P}$, ${\bf U}$
and ${\bf V}$. These are
\bea
a &=& U^2+V^2+ P^2 \nonumber\\
b &=& {\bf P}\cdot {\bf U}\times {\bf V}\nonumber \\
c &=& |{\bf U}\times{\bf V}|^2 +
|{\bf P}\times{\bf U}|^2 + |{\bf P}\times{\bf V}|^2\, .
\eea
An explicit computation shows that
\be
\det\{Q,Q\} = P(H)
\ee
where $P(H)$ is the quartic polynomial
\be
P(H)=H^4 - 2aH^2 - 8bH +  a^2 - 4c \label{quartic}
\ee
The fact that $\{Q,Q\}$ is a positive real symmetric matrix imposes a
bound on $H$ in terms of the invariants $a,b,c$. Specifically,
\be
H\ge E(a,b,c)
\ee
where $E(a,b,c)$
is the largest root of $P(H)=(H-\lambda_1)(H-\lambda _2)(H-\lambda
_3)(H-\lambda _4)$. Since the sum of the roots vanishes, the largest
root $E$ is necessarily non-negative. The number of supersymmetries
preserved is then the number of roots equal to $E$. The vacuum
configuration has all roots equal with $E=0$. In all other cases $E$ is
strictly positive and the number of roots equal to it is 1,2 or 3,
corresponding to 1/4,1/2 or 3/4 supersymmetry. 

Our first task, to be undertaken below, is to analyse the conditions required
for the realization of each of these possibilities. We will then show how the
stability of states preserving supersymmetry, alias `BPS states', is guaranteed
by the supersymmetry algebra. Although all model-independent consequences of
supersymmetry are encoded in the supersymmetry algebra, the extraction of these
consequences for BPS states is facilitated by methods that involve only the
constraints on the Killing spinors associated with these states, and we show in
the subsequent subsection how these methods can be used to learn about
restrictions imposed by the preservation of  supersymmetry on intersecting domain
walls. We conclude with a discussion of 3/4 supersymmetry, and a proof that this
fraction is not realized in the WZ model.

\subsection{Supersymmetry fractions}

The analysis of the conditions on the invariants $a,b,c$ required for the
preservation of the various possible fractions of supersymmetry is fairly
straightforward when the polynomial $P(H)$ has at least two equal roots, and
is especially simple when there are three equal roots. We shall therefore
begin with the case of three equal roots, followed by the case of two equal
roots, arriving finally at the generic case. 

The quartic polynomial $P(H)$ has three equal roots if
\be
c={a^2\over 3}, \qquad b=\mp({a\over 3})^{3/2}\, ,
\ee
and the roots are
\be
\lambda_1=\lambda_2=\lambda_3\equiv \lambda=\pm ({a\over 3})^{1/2},\qquad
\lambda_4=-3\lambda\, .
\ee
If $\lambda$ is positive then we have the BPS bound $H\ge \lambda$, and
charge configurations saturating this bound preserve 3/4 supersymmetry. If
$\lambda$ is negative then we instead find the BPS bound $H\ge -3\lambda$, with
only 1/4 supersymmetry being preserved by charge configurations that saturate
it.

Charge configurations preserving 1/2 supersymmetry can occur only when
$P(H)$ has two equal roots. The conditions for the special case in which
$\lambda_1=\lambda_2$ and $\lambda_3=\lambda_4$ are
\bea
b=c=0\nn
\lambda_1=-\lambda_3=\pm {\sqrt a}
\eea
In the  more general case when $\lambda_1=\lambda_2\equiv\lambda$ and
$\lambda_3\equiv \rho$ we have $\lambda_4=-(2\lambda +\rho)$.
If $\lambda=0$ we have $a^2=4 c$, $b=0$ and $\rho^2=2 a$, with 1/4
supersymmetry when $H=|\rho|$. Otherwise we find the condition
\be
4 a^3 b^2 +27 b^4 -18 a b^2 c -a^2 c^2 +4 c^3 =0
\ee
with
\be
3\lambda^2 = a\pm 2 (a^2-3 c)^{1/2}\, ,\qquad
\rho^2+2 \lambda \rho + 3 \lambda^2 = 2a\, .
\ee
with 1/2 supersymmetry possible when $\lambda$ is the largest root.

The general case of four unequal roots is quite complicated, unless $b=0$, in
which case
\be \label{roots}
(\lambda_1, \lambda_2 ,\lambda _3, \lambda _4)=
\bigl ( \sqrt{a+2\sqrt{c}} \thinspace,
\sqrt{a-2\sqrt{c}}\thinspace ,-\sqrt{a-2\sqrt{c}} \thinspace
,-\sqrt{a+2\sqrt{c}} \thinspace)\, .
\ee
One way to achieve $b=0$ is to set ${\bf P}={\bf 0}$. In this case the bound on
$H$ becomes
\be\label{satbound}
H \ge \sqrt{U^2+V^2 + 2|{\bf U}\times{\bf V}|}\, .
\ee
Note that this becomes $H \ge  |{\bf U}| + |{\bf V}|$ when ${\bf U}\cdot{\bf
V}={\bf 0}$, which is typical of 1/4 supersymmetric orthogonal intersections of
branes. The four eigenvalues of $\{Q,Q\}$ are, in
order of increasing magnitude,
\be
H - \sqrt{a+2\sqrt{c}},\qquad  H - \sqrt{a-2\sqrt{c}},\qquad
H + \sqrt{a-2\sqrt{c}},\qquad  H + \sqrt{a+2\sqrt{c}}\, .
\ee
The first of these vanishes when the bound is saturated. The last two are never
zero unless all four vanish, which is the vacuum charge sector. The second
eigenvalue equals the first only when $c=0$, so in this case there are two zero
eigenvalues when the bound is saturated and we have 1/2 supersymmetry.
Otherwise we have 1/4 supersymmetry.

As emphasized earlier, static configurations need not have ${\bf P}={\bf 0}$
because ${\bf P}$ may have an interpretation as a domain-wall junction charge,
rather than 3-momentum (in general it must be interpreted as a sum of the
3-momentum and a string junction charge). Nevertheless, one may still have
$b=0$ if ${\bf U}\times {\bf V}$ vanishes, which it will do if, say, ${\bf
V}=0$. In this case, the results are exactly as in the ${\bf P}={\bf 0}$ case
just analysed but with ${\bf V}$ replaced by ${\bf P}$. In particular, if ${\bf
P}\cdot {\bf U}=0$ we then have $H\ge |{\bf P}| + |{\bf U}|$, and static 1/4
supersymmetric configurations have $H\ge |{\bf P}| + |{\bf U}|$. For this case,
we can bring the charges to the form
\be
{\bf P}= (0,0,Q) \, ,\qquad {\bf U}=({u_1,u_2,0})\, ,\qquad {\bf V}=(0,0,0)\, .
\ee
where $Q$ is a junction charge.
This case is the one analysed in \cite{GT}, with $T=u_1+iu_2$ being the complex
scalar charge in the D=3 supersymmetry algebra obtained by dimensional
reduction on the 3-direction. In agreement with \cite{GT} we find that
$H= |T|+|Q|$, so the junction charge contributes positively to the energy of
the whole configuration.

More generally, we might have
\be
{\bf P}=(0,0,Q)\, ,\qquad {\bf U}=(u_1,u_2,0)\, ,\qquad {\bf V}=(v_1,v_2,0)\, .
\ee
This case was analysed in \cite{oda}, and an explicit realization of it was
found in a model with several chiral superfields; in this model the charge Q is
again associated with a domain wall junction. In agreement with
\cite{oda} we find the four roots to be
\bea
\lambda_1 &=& -Q + \sqrt{(u_2+v_1)^2 + (u_1-v_2)^2}\nonumber\\
\lambda_2 &=& -Q - \sqrt{(u_2+v_1)^2 + (u_1-v_2)^2}\nonumber\\
\lambda_3 &=& Q - \sqrt{(u_2-v_1)^2 + (u_1+v_2)^2}\nonumber\\
\lambda_4 &=& Q + \sqrt{(u_2-v_1)^2 + (u_1+v_2)^2}\, .
\eea
Note that the four roots are distinct, in general, and (in contrast to
the previous case) $b\ne0$. If $Q$ is positive and $\lambda_1$
is the largest root, the junction charge $Q$ contributes negatively to
the total energy as in \cite{oda}.

The case just considered is a special case of the larger
class of configurations
with $b\ne0$ for which $P(H)$ has four distinct roots. At this point the
analysis becomes quite complicated, and we shall not pursue it further.

\subsection{Stability of BPS states}

Our aim in this subsection is to prove the stability of BPS
states. We begin by considering the possible
decay of a general state, not necessarily BPS,
with energy $H_3$ into two
other states, not necessarily BPS, with energies  
$H_1$ and $H_2$. This can be
represented schematically as
\be\label{fission}
(state)_3 \rightarrow (state)_1 + (state)_2\, .
\ee
Let us write
\be
\{Q,Q\} = H + K(a,b,c)\, ,
\ee
where $K$ is a traceless symmetric matrix, and $(a,b,c)$ are the three
$SU(2)\times SU(2)_R$ invariants introduced previously.  
Conservation of charges and energy requires that
\bea
H_3&=& H_1+H_2\\
K_3&=& K_1 + K_2
\eea
where $K_i= K(a_i,b_i,c_i)$, with $(a_i,b_i,c_i)$ being the values of the
invariants $(a,b,c)$ for the $i$th state. 
Since the matrices $H_i + K_i$ are positive definite they are subject
to the Minkowski reverse triangle inequality (see e.g. \cite{hardy})
\be\label{hardy}
[\det (H_3+K_3)]^{1\over 4} \ge
[\det(H_1 + K_1)]^{1\over4} + [\det(H_2 + K_2)]^{1\over4}\, .
\ee

We now want to see the consequences of supposing state 3 to be BPS.
We observe that the left hand side of (\ref{hardy}) vanishes if state 3
is BPS, but the right hand side can
vanish only if both states 1 and 2 are also BPS. The extension to more
than two decay products is immediate so we conclude
that any unstable BPS state would have to decay into other BPS states.

To complete the proof of stability we now show that a BPS state 
cannot decay into other BPS states. 
A BPS state has an energy $H=E(K)\equiv E(a,b,c)$ where $E(K)$ is 
the largest value of $H$ for which
$\det(H+K)=0$. An equivalent characterization of $E(K)$ is as the smallest
eigenvalue of $K$. It follows that $E(K)= min \left(\zeta^T K \zeta\right)$,
where $\zeta$ is a commuting spinor normalized such that $\zeta^T\zeta=1$ but
otherwise arbitrary. From this and the fact that $min (a+b) \le min (a) +
min(b)$, we deduce the triangle 
inequality
\be\label{tri}
E(K_1+K_2) \le E(K_1) + E(K_2)\, .
\ee
Generic models will have a spectrum of BPS states for which this 
inequality is never saturated. In such cases BPS states are absolutely 
stable. In those cases for which there are BPS energies saturating 
the inequality (\ref{tri}) there may be states of marginal 
stability\footnote{It is well known that marginal stability is the 
mechanism by which BPS states `decay' as one moves in the space of 
parameters defining certain theories, but this is a discontinuity of the 
BPS spectrum as a function of parameters and not a process within 
a given theory.}. The inequality (\ref{tri}) is saturated
when $K_1$ and $K_2$ are proportional, with positive constant of
proportionality, but this is only a sufficient condition for equality. 
Another sufficient condition, which we believe to be necessary, is the 
coincidence, up to normalization, of
the eigenvectors of $K_1$ and $K_2$ with lowest eigenvalue.

It is instructive to see how the above comments apply to the special case 
in which $H+K=C\gamma^\mu P_\mu$. The Minkowski inequality becomes
\be
\sqrt{-(P_1 + P_2)^2} \ge \sqrt{-P_1^2} +\sqrt{-P_2^2}\, .
\ee
Since $\sqrt{-P^2}$ is the rest mass $m$ of a particle with
4-momentum $P$, we learn that
\be\label{reverse}
m_3\ge m_1 + m_2\, .
\ee
This is the familiar rule that the sum of the masses of the decay products
cannot exceed the mass of the particle undergoing decay. Given that $m_3=0$ we
deduce that $m_1=m_2=0$, so if a massless particle decays into two other
particles those two particles must also be massless.
For this special case the triangle inequality (\ref{tri}) reduces to
\be\label{obvious}
|{\bf P}_1 + {\bf P}_2| \le |{\bf P}_1| + |{\bf P}_2|\, , 
\ee
which is saturated if and only if ${\bf P}_1$ and ${\bf P}_2$ are
parallel, and in this case there is no phase space for the decay.

\subsection{Domain Walls at Angles}

Each supersymmetric configuration is associated with a set of Killing spinors
$\epsilon$ which span the kernel of $\{Q,Q\}$. With the
exception of the vacuum configuration, these spinors are subject to constraints
that reduce the dimension of the space that they span. Some properties of
supersymmetric configurations follow directly from the nature of these
constraints. In particular, intersecting brane configurations can be considered
as configurations obtained from parallel branes by rotation of one or more of
them. The constraints can be similarly obtained, and then analysed to determine
the dimension of the space of Killing spinors they allow \cite{BDL}. We shall
apply this analysis here to intersecting domain walls of N=1 D=4 theories.

We begin with two coincident domain walls, corresponding to the constraint
\be\label{firstcon}
\gamma_{013}\epsilon=\epsilon\, .
\ee
We then rotate one of them  around the
3-axis until it makes an angle $\beta$ in the 12-plane, and simultaneously
rotate by some angle $\alpha$ in the electric-magnetic charge space. This
operation is represented by the matrix
\be
R= e^{{1\over2}\alpha\gamma_5}e^{{1\over2}\beta\gamma_{12}}\, ,
\ee
which satisfies
\be\label{gary}
\gamma_{013}R^{-1} = R\gamma_{013}\, .
\ee
The constraint on the Killing spinor $\epsilon$ imposed by
the rotated brane is
\be\label{secondcon}
R\gamma_{013}R^{-1}\epsilon = \epsilon\, .
\ee
Using (\ref{gary}) and (\ref{firstcon}), one easily verifies that this
second constraint is equivalent to
\be\label{thirdcon}
\left(R^2-1\right)\epsilon=0\, .
\ee
It is not difficult to show that this equation has no non-zero solutions for
$\epsilon$ unless $\alpha \pm\beta =0$.
We thus have
\be
\label{dgvkjfghsk}
R= e^{\alpha\Sigma}\, ,\qquad \qquad
\Sigma= {1\over2}\left(\gamma_5 \pm \gamma_{12}\right)\, .
\ee
Using the identity $\Sigma^3=-\Sigma$ one can establish that
\be
R^2-1 = (2R)(\sin\alpha \,\Sigma) \, .
\ee
Since $2R$ is invertible, it follows that (\ref{thirdcon}) is equivalent to
\be
\sin\alpha\, \Sigma\, \epsilon=0\, .
\ee
This is trivially satisfied if $\sin\alpha=0$. Otherwise it reduces to
$\Sigma\epsilon=0$, which is equivalent to
\be\label{pcon}
\gamma_{03}\epsilon=\pm\epsilon\, .
\ee
If this is combined with (\ref{firstcon}) we deduce that
\be\label{altcon}
\gamma_5\gamma_{023}\epsilon= \mp\epsilon\, ,
\ee
which is the constraint associated with a purely magnetic domain wall in the
23-plane. We may take any two of these three constraints as the
independent ones; the choice (\ref{firstcon}) and (\ref{altcon}) have an
obvious interpretation as the constraints associated with the orthogonal
intersection of an electric wall with a magnetic one. This constitutes the
special $\alpha=\pi/2$ case of the more general configuration of
rotated intersecting branes that we have been studying. But we have now derived
these constraints for {\sl any} angle $\alpha\ne 0,\pi$.
The fraction of supersymmetry preserved by the general rotated brane
configuration is therefore the same as the fraction preserved in the special
case of orthogonal intersection. Standard arguments can now be used to show
that this fraction is 1/4.

We have thus shown that starting from a 1/2 supersymmetric configuration of two
parallel coincident domain walls with normal ${\bf n}$, one of them may be
rotated relative to the other by an arbitrary angle in a plane containing ${\bf
n}$, preserving 1/4 supersymmetry, provided that the charge of the rotated
wall is simultaneously rotated by the same angle in the `electric-magnetic'
charge space. In practice it may not be possible for the domain walls
to intersect at arbitrary angles (preserving supersymmetry). For
example, in the $\bZ_3$-invariant model discussed in \cite{GT},
supersymmetric intersections are necessarily at $2\pi/3$ angles.
But such additional restrictions are model-dependent. What we learn
from the supersymmetry algebra is the model-independent result that
{\sl the angle  separating 1/4 supersymmetric intersecting
domain walls must equal the
angle between them in the `electric/magnetic' charge space}.

Since the constraint (\ref{pcon}) is associated with non-zero $P_3$
we also learn from the above analysis that we can include this charge,
provided it has the appropriate sign, which is
determined by the sign in (\ref{dgvkjfghsk}),
without affecting the constraints imposed by 1/4 supersymmetry,
although we then leave the class of configurations for which $b=0$.
Setting $P_3\ne0$ might be considered as performing a boost along
the 3-direction except for
the previously noted fact that $P_3$ is not necessarily to be
interpreted as momentum. Nevertheless, as a terminological convenience
we shall {\sl call} ${\bf P}$ the `3-momentum' in what follows.
Consider the charge configuration obtained by adding
the charges of an electric brane in the 13-plane with a brane rotated in the
12-plane, preserving 1/4 supersymmetry, and then adding momentum in the 3
direction:
\bea \label{mixed}
{\bf U} &=& v\cos\alpha(\sin\alpha,-\cos\alpha,0)+ (0,-u,0) \nonumber \\
{\bf V} &=& v\sin\alpha \left (\sin\alpha, -\cos\alpha,0\right) \nonumber\\
{\bf P} &=& (0,0,p)
\eea
We now have
\bea
a&=&u^2+v^2+2 uv \cos^2\alpha+p^2\nn
b&=&puv\sin^2\alpha\nn
c&=&u^2v^2\sin^4\alpha +p^2(u^2+v^2+2 u v \cos^2\alpha)
\eea
One can show that the eigenvalues of $\{Q,Q\}$ are
\be
H + p \pm \sqrt{u^2+v^2+2uv\cos 2\alpha}\, ,\qquad H-p\pm (u+v)
\label{eigs}
\ee
For $u,v,p\ge0$, we conclude that $H\ge p+u+v$ and that 1/4 supersymmetry is
preserved when the bound is saturated. Note that in this case
\be
\{Q,Q\} = u\left(1-\gamma_{013}\right) + v\left(1-\gamma_{013}R^2\right) +
p\left(1-\gamma_{03}\right)\, ,
\ee
for the upper sign in (\ref{dgvkjfghsk}),
confirming that the projections remain unchanged by the inclusion of momentum.

\subsection{3/4 Supersymmetry}

Continuing the above analysis, we now turn to the case in which $u,v,p$
are not necessarily all positive because this case includes the
possibility of domain wall configurations preserving 3/4 supersymmetry
\cite{GH}. Consider the case $\alpha= \pi/2$ for an electric wall and a magnetic
wall intersecting at right angles, so that the eigenvalues (\ref{eigs}) are
\be\label{eigen}
H + p \pm (u-v)\, ,\qquad H-p\pm (u+v)\, .
\ee
It follows that $H$ is bounded below by each of the eigenvalues
\bea
\lambda_1&=&p-u-v\nn
\lambda_2&=&v-u-p\nn
\lambda_3&=&u-v-p\nn
\lambda_4&=&u+v+p\, .
\label{uvp}
\eea
If only one of the charges is non-zero, $u$ say, then we obtain the standard
BPS bound, $H\ge |u|$, which is saturated by the electrically charged BPS
domain wall.  With two charges, $u$ and $v$ say, we obtain $H \ge |u+v|$ and
$H\ge |u-v|$, and when the stronger of these is saturated we have the
intersecting domain wall configuration preserving 1/4 supersymmetry. With all
three charges, there are four bounds corresponding to the four eigenvalues and
1/4 supersymmetry is preserved, generically, when the strongest bound is
saturated. There are then two subcases to consider according to whether or not
$\lambda_4$ is the largest eigenvalue. If $\lambda_4$ {\sl is} the largest
eigenvalue,  as happens, for example, when $u,v,p$ are all positive, then we
recover the standard 1/4 supersymmetric case considered above, unless two of
the
three charges $u,v,p$ vanish in which case 1/2 supersymmetry is preserved. If
$\lambda_4$ is {\sl not} the largest eigenvalue then one of the others is, and
we may choose it to be $\lambda_1$ because the other possibilities are
related to this one by $SU(2)_R$ transformations. Given this, $H$ is bounded
below by $p-u-v$ and if there is a state saturating this bound with $H=p-u-v$
then the eigenvalues of $\{Q,Q\}$ are
\be
0\, ,\qquad 2(p-v)\, ,\qquad 2(p-u)\, ,\qquad -2(u+v)\, .
\ee
It follows that 1/4 supersymmetry is preserved {\sl generically} but more
supersymmetry is preserved for special values of the charges. The possibility
of this kind of enhancement of supersymmetry, including the possibility of 3/4
supersymmetry, was recently discussed in \cite{GH} and the case under
consideration here is very similar. If $p=v$ or $p=u$ or $u=-v$, then a charge
configuration saturating the BPS bound will preserve 1/2 supersymmetry and if
$p=u=v$ or $u=-v=\pm p$ then 3/4 supersymmetry will be preserved. Thus, a
charge configuration  saturating the bound $H \ge  \lambda_1$ will preserve 1/4
supersymmetry for {\sl generic} values of the charges, but 1/2 or 3/4
supersymmetry for certain special values.

We should stress that the above analysis is purely algebraic and it is an open
question whether there exists a physical model with domain wall configurations
preserving 3/4 supersymmetry. As we now show, this possibility is not realized
by the WZ model.

\subsection{BPS Solutions of the Wess-Zumino Model}

The WZ model is known to admit both 1/4 and 1/2 supersymmetric
classical solutions, which (at least potentially) correspond to states in the
quantum theory. We shall show here that there are no classical solutions
preserving 3/4 supersymmetry. We shall begin by considering purely bosonic 
field configurations and then extend the result to arbitrary classical
configurations.

The fields of the WZ model belong to a single chiral superfield,
the components of which are a complex physical scalar
$\phi=A+iB$, a complex two-component spinor, which is equivalent to a
4-component Majorana spinor $\lambda$, and a complex auxiliary field
$F=f+ig$. We will continue to use a real representation of the four Dirac
matrices $\gamma^\mu$. For purely bosonic field configurations we need only
consider fermion supersymmetry transformations. Our starting point will
therefore be the (off-shell) supersymmetry
transformation of $\lambda$, which takes the form $\delta\lambda = M\epsilon$,
where $\epsilon$ is a real constant spinor parameter and $M$ is the real
$4\times 4$ matrix
\be\label{MMM}
M= \gamma^\mu (\partial_\mu A +\gamma_5 \partial_\mu B) + f +
\gamma_5 g\, .
\ee
This transformation is valid for the spinor component of any
chiral superfield.
The Wess-Zumino model is characterised by the fact that the auxiliary field
equation is
\be
F\equiv f+ig = W'(\phi)\, ,
\ee
where $W'(\phi)$ is the derivative with respect to $\phi$ of the holomorphic
superpotential $W(\phi)$.

A bosonic field configuration of the WZ model will be supersymmetric if there
is a spinor field $\epsilon$ that is both annihilated by $M(x)$, for all $x$,
and covariantly constant with respect to a metric connection on $\bE^{(1,3)}$.
Thus, for there to be
$n$ preserved supersymmetries it is a necessary condition
that $M(x)$ has an $n$  dimensional kernel for each $x$.
Our strategy for showing that there are
no 3/4 supersymmetric field configurations will be to
analyse necessary conditions for the matrix $M_0\equiv M(x_0)$ at a fixed
point $x_0$ to have an $n$-dimensional kernel.

We begin by noting that a WZ field configuration
can preserve 1/4 supersymmetry only if
$\det M_0$ vanishes, which is equivalent to
\be\label{det}
\left[(\partial A)^2 + (\partial B)^2 -f^2 -g^2\right]^2 =
4\left[ (\partial A)^2(\partial B)^2 -
(\partial A \cdot \partial B)^2\right]\, .
\ee
This condition is necessary for the preservation of at least 1/4
supersymmetry in any model with a single chiral
superfield, and in particular in the WZ model.
Configurations preserving more than 1/4 supersymmetry are characterized by
additional constraints on the fields. Necessary constraints can be
found very easily by making use of the fact that $M_0$ can be brought to (real)
upper-triangular form by a  similarity transformation. 
We may therefore assume that $M_0$ is upper triangular. 
If, in addition, it has a 2-dimensional kernel then
it may be brought to the form
\be
\pmatrix{0&0&*&*\cr &0&*&*\cr & &*&*\cr & & &*}
\ee
where $*$ indicates an entry that is not zero (or not necessarily zero
in the case of the off-diagonal entries).
This matrix has the property that
\be\label{half}
2\tr M_0^3 -3\tr M_0\, \tr M_0^2 + (\tr M_0)^3 =0\, ,
\ee
and substituting (\ref{MMM}) we learn that
\be\label{halftwo}
f\left[f^2 + g^2 -(\partial A)^2 - (\partial B)^2\right]=0\, .
\ee
This condition is therefore necessary for a field configuration to preserve 1/2
supersymmetry.

Similarly, any upper-triangular matrix with a
3-dimensional kernel can be brought to the form
\be
\pmatrix{0&0&0&*\cr &0&0&*\cr & &0&*\cr & & &*}
\ee
This matrix satisfies both (\ref{half}) and
$\tr M_0^2=(\tr M_0)^2$, in addition
to (\ref{det}). These conditions, which are therefore necessary for 3/4
supersymmetry, are equivalent to the joint conditions
\bea\label{joint}
f &=&0 \nonumber\\
g^2 &=& (\partial A)^2 + (\partial B)^2\nonumber\\
(\partial A)^2(\partial B)^2 &=& (\partial A \cdot \partial B)^2\, .
\eea

We are now in a position to show that there are no 3/4 supersymmetric WZ field
configurations (other than the vacuum which has $4/4$ supersymmetry). The
conditions (\ref{joint}) must be satisfied by such a field configuration.
We shall analyse these conditions at a fixed point $x=x_0$ and consider
separately the cases in which $g=0$ and $g\ne0$ at that point. If
$g=0$ then the second condition  in (\ref{joint}) implies that at $x_0$
{\sl either} the 4-vectors $\partial A$ and $\partial B$ are both null {\sl or}
one is spacelike and the other is timelike. The latter option contradicts the
third of eqs (\ref{joint}) so both are null. It then follows from
(\ref{joint}) that $\partial A$ and $\partial B$ are parallel, so that
\be\label{configone}
f=g=0\, , \qquad \partial A = \alpha\,  v \, ,\qquad  \partial B=\beta\,  v
\ee
where $\alpha$ and $\beta$ are constants and $v$ is a null 4-vector. This
field configuration is therefore a candidate for 3/4 supersymmetry, but because
the conditions leading to it were not sufficient for 3/4 supersymmetry this
must be checked. In fact, it is readily shown that the matrix $M$ corresponding
to the configuration (\ref{configone}) has only a two-dimensional kernel
so that at most 1/2 supersymmetry can be preserved.

The remaining candidates for 3/4 supersymmetry in the WZ model arise from
field configurations in which  $f$ vanishes but $g$ is non-zero.
Then (\ref{joint})  implies that at $x_0$ either $\partial A$ and  $\partial B$
are both spacelike, or one is spacelike and the other is null. Suppose first
that
either $\partial A$ or $\partial B$ is null. In the case in which
$\partial B$ is  null we have
\be\label{configthree}
f=0\, \qquad \partial A= g s \, ,\qquad \partial B= v\, ,
\ee
where $v$ is a null vector orthogonal to a spacelike vector $s$ normalized such
that $s^2=1$.  For this configuration one can check that the matrix $M$
generically has a one dimensional kernel, and has a two dimensional kernel when
either $g=0$ or $\beta=0$. The case in which $\partial A$ is null is similar,
with the same result that at most 1/2 of the supersymmetry is preserved.

If neither $\partial A$ nor $\partial B$ is null then they are both spacelike
and we
can arrange for them to take the form
\bea\label{configfour}
\partial B &=& \beta(0,1,0,0)\nonumber\\
\partial A &=& \alpha(\sin\theta,\cos\theta,0,\sin\theta)
\eea
with $g^2= \alpha^2\cos^2\theta + \beta^2$. One then finds that
the kernel of $M(x_0)$ is 2-dimensional if $\alpha\beta\sin\theta=0$ and
otherwise 1-dimensional. Configurations
of the form (\ref{configfour}) can therefore preserve at most 1/2
supersymmetry.

We have now shown that there are no non-vacuum {\sl bosonic} WZ field
configurations that preserve 3/4 supersymmetry. We now wish to consider whether
this remains true when we consider general configurations that are not
necessarily bosonic. This question is perhaps best posed in the context of the
quantum theory, which we will not consider here, but it can also be posed
classically by taking all fields to be supernumbers with a `body' and a
nilpotent `soul'. Any general field configuration of this kind preserving 3/4
supersymmetry must have a body preserving at least 3/4 supersymmetry and, as we
have just seen, the vacuum configuration is the only candidate. It follows that
the only remaining way in which a classical field configuration could be
3/4 supersymmetric is if the 4/4 supersymmetry of the bosonic vacuum
configuration is broken to 3/4 by fermions. Preservation of any
fraction of supersymmetry in a fermionic background requires the
vanishing of the supersymmetry transformations of the bosons. For the
WZ model this implies  ($\bar\lambda \equiv \lambda^TC$)
\be
\bar\lambda\epsilon=0\, ,\qquad
\bar\lambda\gamma_5\epsilon =0\, ,
\ee
and for 3/4 supersymmetry there must be a three-dimensional space of
parameters $\epsilon$ for which this condition holds.
At a given point in space we may choose, without loss of generality, a
basis in spinor space such that $C\epsilon=(0,*,*,*)^T$, where an
asterisk indicates an entry that may be non-zero. The first
equation then implies that $\lambda^T =(*,0,0,0)$ and the
second that $\lambda^T\gamma_5=(*,0,0,0)$. But since $\gamma_5$ is
both real and satisfies $\gamma_5^2=-1$ these conditions are not
mutually compatible. This concludes our proof that the WZ model has no
non-vacuum classical configurations, bosonic or otherwise,
that preserve 3/4 supersymmetry

\section{The geometry of supersymmetry}

We now turn to a discussion of the geometry associated with BPS representations
of the algebra (\ref{susyalg}), which we may re-write in terms of a
positive semi-definite symmetric bispinor $Z$ as $\{Q,Q\}=Z$. The positivity of
$\{Q,Q\}$ implies that $Z$ is a vector in a convex cone, with the boundary of
the cone corresponding to the BPS condition $\det Z=0$. We shall first explain
some of the geometry associated with convex cones, and how it relates to BPS
states. We will then explain how this ties in with the theory of Jordan
algebras.

\subsection{Convex cones}

Let us begin with the standard D=4 N=1 supersymmetry algebra, in which
case $Z=\gamma\cdot P$ and the positivity of $\{Q,Q\}$
implies that $P$ lies {\sl either} in the
forward lightcone of D=4 Minkowski momentum-spacetime {\sl or} on its
boundary, the lightfront. In the latter case, $P^2=0$ and any states
with this 4-momentum are BPS, preserving 1/2 supersymmetry.
The forward lightcone in momentum space and the forward
lightcone in position space are both  examples of convex
cones. An $n$-dimensional cone ${\cal C}$ is a subspace of an
$n$-dimensional vector space $V$ with the property that $\lambda x \in {\cal
C}$
for all $x\in {\cal C}$ and all real positive $\lambda$. The cone is convex if
the sum of any two vectors in the cone is also in it. The dual cone is then
defined as follows. Let $y$ be a vector in the dual vector space $V^*$ and
let $y\cdot x$ be a bilinear map from $V\times V^*$ to
$\bR$. The dual cone ${\cal C}^*$ is the subspace of $V^*$ for
which $y\cdot x >0$ for all $x\in {\cal C}$.

Given a translation-invariant measure on $V$ we can associate with
each convex cone in $V$ a characteristic function $\omega$ defined by
\be
\omega^{-1}(x) = \int_{{\cal C}^*} e^{-y\cdot x} d^n y\, .
\ee
As all translation-invariant measures are multiples of any given
translation-invariant measure, this formula defines $\omega$ up to a
scale factor, but this ambiguity will not affect the statements to
follow. 
The cone is foliated by hypersurfaces of constant $\omega$, with the
limiting hypersurface $\omega=0$ being the boundary of the cone.
In the case of the forward light cone in D=4 Minkowski spacetime the vector
space $V$ is $\bR^4$ and $\omega={\cal N}^2$, where ${\cal N}(x)= -\eta _{\mu
\nu } x^\mu x^\nu$ is the quadratic form defined by the Minkowski
metric $\eta$ (we adopt a `mostly plus' metric convention). The
hypersurfaces of constant
$\omega$ are therefore  hyperboloids homothetic to $SO(1,3)/SO(3)$. Note that
this is a symmetric space; this is a general feature of self-dual homogeneous
convex cones, of which the forward lightcone in Minkowski space is an example.
Homogeneous convex cones that are not self-dual are foliated by homogeneous
spaces that are not symmetric spaces. 

Because, in this example, $\omega$ is determined by a {\sl quadratic} function
${\cal N}$, the  vector space $V=\bR^4$ can be viewed as a metric space, with
Minkowski metric $\eta$. More generally, $\omega$ is not quadratic and hence
does not furnish $V$ with a metric. Nevertheless, $\omega$ does provide a
positive definite metric for ${\cal C}$ (obviously, this differs from the
Minkowski metric of the `quadratic' case discussed above). Let us first note
that, by the definition of a cone, the map  $D: x\mapsto
\lambda x$ is an automorphism, in that $Dx \in {\cal C}$ if $x\in
{\cal C}$. It follows immediately that $\omega(x)$ is a homogeneous function of
degree $n$. A corollary of this is that $\pi (x)\cdot x= 1$ where
\be\label{HJ}
\pi(x)= {1\over n} {\partial \log \omega\over \partial x}\, .
\ee
Thus, $\pi\in {\cal C}^*$, and as $x$ ranges over all vectors in ${\cal C}$ so
$\pi$ ranges over all vectors in ${\cal C}^*$. One can now introduce a metric
$g$ on ${\cal C}$ with components\footnote{For the forward light-cone in
Minkowski spacetime with Minkowski metric $\eta$, we have
$g_{ij}= (x^2)^{-2}( 2 x_i x_j-x^2\eta_{ij})$ where $x^2=
\eta_{ij}x^ix^j$ and $x_i=\eta_{ij}x^j$, so that $\pi _i = (x^2)^{-1} x_i$.}
\be\label{posmet}
g_{ij} = -{1\over n}\partial_i\partial_j \log\omega (x)\, .
\ee
One may verify that
\be
\pi_j = x^ig_{ij}\, .
\ee
The map from ${\cal C}$ to ${\cal C}^*$ provided by the metric (\ref{posmet})
has a natural interpretation in terms of
Hamilton-Jacobi theory: if $\log\omega$ is interpreted as a
characteristic function in the sense of Hamilton, then $\pi$ as defined by
(\ref{HJ}) is the conjugate momentum.

A feature of the metric $g$ is that it is invariant under
automorphisms of ${\cal C}$. For example it follows from the
homogeneity of $\omega$ that the linear map $D$
is an isometry of $g$. The group of automorphisms will generally be
a semi-direct product of $D$ with some group $G$ that acts on the leaves of the
foliation. The cone is homogeneous if $G$ acts transitively. A homogeneous cone
is foliated by homogeneous hypersurfaces of the form $G/H$ for some
isotropy group $H$. For a self-dual cone this homogeneous space is also a
symmetric space. As already mentioned, the forward light cone in $\bE^{(1,3)}$
is foliated by hyperboloids homothetic to $SO(1,3)/SO(3)$, so
$G$ is the (proper orthochronous) Lorentz group. The metric induced on
each leaf of the foliation by the metric $g_{ij}$ of the cone is the
positive-definite $SO(1,3)$-invariant metric on $SO(1,3)/SO(3)$.

Let us now turn to the general D=4 N=1 supersymmetry algebra
$\{Q,Q\}=Z$. The bispinor charge $Z$ can be interpreted as a vector in
the convex cone of positive-definite real $4\times4$ symmetric
matrices. This is a cone in $\bR^{10}$ which, since $Z$ includes the
4-momentum, we may consider as a `momentum-space' cone ${\cal C}^*$. We
set aside to the following subsection consideration of the corresponding
`position
space' cone ${\cal C}$. The characteristic function of ${\cal C}^*$ 
is\footnote{Note that
$\omega^2$ is a polynomial. A theorem of Koecher states that
$\omega^2$ is a polynomial for all self-dual homogeneous convex
cones.}
\be
\omega(Z)=(\det Z)^{5\over2}\, .
\ee
The cone is again a self-dual homogeneous one, and is foliated by
symmetric spaces that are homothetic to $SL(4;\bR)/[SO(4)]$. 
Of principal interest here is the boundary of ${\cal C}^*$, defined by
$\det Z=0$, because this is the condition for
preservation of supersymmetry. The geometry of this boundary is
now rather more complicated than it was before.

The basic observation required to understand this geometry is that
the cone is a stratified space with  strata ${\cal S}_n$, $n=0,1,2,3,4$, where
${\cal S}_n$ is
the  subspace in which  at least $n$  of the four
eigenvalues vanish, corresponding to at least $n$ supersymmetries being
preserved, and ${\cal S}_{n+1} $ is the boundary of ${\cal S}_n$.
The boundary of the cone is the space ${\cal S}_1$, which is
the 9-dimensional space of
matrices of rank 3 or less.
The boundary of this is the space ${\cal S}_2$ of matrices of rank 2 or less
which
make up a 7 dimensional space. To see why it is 7 dimensional recall that to
specify a matrix of rank 2 it suffices to give the
normalised  eigenvectors with non-vanishing eigenvalues together with  their
eigenvalues. The two eigenvectors define a 2-plane in
$\bR ^4$, corresponding to an element of the 4-dimensional Grassmannian
$SO(4)/(SO(2)\times SO(2))$. Giving the orientation of the eigenvectors within
the 2-plane means specifying one of the
$SO(2)$ factors. In other words the basis of
2 eigenvectors corresponds to the 5
dimensional Stiefel manifold $SO(4)/SO(2)$. Taking into account the
two eigenvalues we have a 7-dimensional space, as claimed.
The boundary of this stratum is the set ${\cal S}_3$ of matrices of rank 1 or
less. These span a 4-dimensional space, since a rank 1 matrix is specified by 
the direction, up to a sign, of its eigenvector with non-zero eigenvalue 
together with the eigenvalue. This is a point in $\bR P^3\times \bR^+$.
Finally, the boundary of ${\cal S}_3$ is the stratum ${\cal S}_4$ 
consisting of the zero matrix, which is the vertex of the cone. 

\subsection{Reverse triangle inequalities}

The Minkowski inequality that we used previously to establish the stability of
BPS states is a special case of a reverse-triangle inequality valid for all
convex cones. Let us define the `length' of a vector in an n-dimensional convex
cone with characteristic function $\omega$ as
\be
L(x) = \omega^{1/n}(x)\, .
\ee
This is a homogeneous function of degree 1. Because the hypersurfaces of
constant $\omega$ are {\sl concave}, this `length' satisfies the {\sl reverse}
triangle inequality
\ben
L(x+x')\ge L(x) + L(x')\, .
\een
with equality if and only if $x$ and $x'$ are proportional.
In the case of the cone of $m\times m$ positive definite hermitian matrices
we have $L(x) = (\det x)^{1/m}$ and the reverse triangle inequality is
the Minkowski inequality
\be
[\det (x+y)]^{1\over m} \ge [\det x]^{1\over m} + [\det y]^{1\over m}\, ,
\ee
with equality if the two matrices are proportional. In the special case
of diagonal matrices, the cone becomes the positive orthant
${\Bbb R}_+^m$ in $\bE^m$. The length of a vector $x=diag (x_1,\dots,x_m)$ in
${\Bbb R}_+^m$ is $L(x)= (x_1\dots x_n)^{ 1/n}$, and Minkowski's inequality for
positive definite matrices reduces to a form of Holder's inequality (see e.g.
\cite{hardy}). The metric $g$ on ${\Bbb R}_+^m$ is the flat metric
$dl^2= (1/n)\sum (d\log x^i)^2$. The automorphism group is the permutation
group $S_m$, which is clearly an invariance of the length.

\subsection{Conformal Invariance}

For the standard D=4 N=1 supersymmetry algebra without central charges
all BPS states have $P^2=0$.
This is the momentum space version of the massless wave-equation, which is
invariant under the action of the conformal group $SU(2,2)$ on compactified
Minkowski spacetime. Our aim here is to show how this generalizes when the
domain wall charges are included. This will turn out to be a straightforward
extension of the standard case, appropriately formulated, so we consider that
first.

It is convenient to identify a point in Minkowski spacetime with
a matrix $X=X^\mu\sigma_\mu$, where $\sigma_\mu=(1,\sigma_1,\sigma_2,\sigma_3)$
are the $2\times 2$ Hermitian sigma-matrices. The conjugate momentum $P$
is then similarly a $2\times 2$ Hermitian matrix and $-P^2$ becomes $\det P$.
(The momentum $P$ should not be confused with the dual variable $\pi$
introduced in the previous subsection.)
Let us now consider the massless particle action
\be
I = \int \, [{\rm tr} PdX -e \det P]
\ee
where $e$ (the einbein) is a Lagrange multiplier for the mass-shell constraint
$\det P=0$. The conformal group $SU(2,2)$ acts on the compactification
of Minkowski space via the fractional linear
transformation
\be\label{fraclin}
X \rightarrow X' = (AX + B)(CX+D)^{-1}
\ee
where the hermiticity of $X'$ requires that
\be
\pmatrix{A&B\cr C&D} \in SU(2,2)\, .
\ee
This implies that
\be\label{pingu}
dX'(CX+D) = (A-X'C)dX\, .
\ee
We deduce from this that the $PdX$ part of the action $I$ is invariant (up to a
surface term) if
\be
P \rightarrow P' = (CX+D)P (A-X'C)^{-1}\, .
\ee
This transformation implies
\be
\det P \rightarrow \det P' = \Omega^{-1} \det P
\ee
where
\be\label{omega}
\Omega = {\det (A-X'C)\over \det (CX+D)}\, .
\ee
The action $I$ is therefore invariant if we assign to the einbein the
transformation $e\rightarrow e'=\Omega e$.

We now wish to determine the analogous symmetry group of the more general
BPS condition $\det Z=0$. The matrix $Z$ can be viewed as a vector in a
10-dimensional vector space. Let $X$ be coordinates of the dual space and
consider the particle  action
\be
I = \int \, [{\rm tr} ZdX - e \det Z]\, .
\ee
Special cases of actions of this type were considered previously by
Cederwall \cite{Ced}, with a motivation derived from Jordan algebra
considerations that we shall explain in the following subsection
(see also \cite{sez,Bandos}).
Now consider the fractional linear transformation
\be\label{fraclingen}
X \rightarrow X' = (AX + B)(CX+D)^{-1}\, ,
\ee
which acts on the compactification of the space of symmetric matrices
\cite{hua}.
The matrix $X'$ will also be real and symmetric provided that
\be
\pmatrix{A&B\cr C&D} \in Sp(8;\bR)\, .
\ee
That is,
\be
A^TD-C^TB=1\, ,\qquad A^TC=C^TA\, ,\qquad B^TD =D^TB\, .
\ee
As before, we deduce (\ref{pingu}) and
from this that the $ZdX$ term is invariant up to a surface
term if
\be
Z \rightarrow Z' = (CX+D)Z (A-X'C)^{-1}\, .
\ee
This implies
\be
\det Z\rightarrow \det Z' = \Omega^{-1} \det Z
\ee
where $\Omega$ has form of (\ref{omega}). We may again take $e\rightarrow e'
=\Omega e$ to achieve an invariance of the action $I$. In this case, the
invariance group is $Sp(8;\bR)$.

Note that this conclusion rests on an interpretation of the
4-dimensional compactified Minkowski spacetime as a subspace of a
ten-dimensional vector space of the $4\times 4$ real
symmetric matrices $X$. A field theory realization of Sp(8;R) would
require fields defined on this larger space. For example, the analogue
of the massless wave equation on Minkowski space is the fourth-order equation
\be
\det (-i\partial /\partial X)\, \Psi =0
\ee
The symmetry group of this equation is $Sp(8;\bR)$. By analogy with the
Minkowski case, we expect this to be the maximal symmetry group of this
equation.

\subsection{Jordan algebras}

The results of the previous subsections have an interpretation in terms of
Jordan algebras.  A Jordan algebra $J$ of dimension $n$ and degree $\nu$ is an
$n$-dimensional real vector
space with a commutative, power associative, bilinear
product, and a norm ${\cal N}$ that is a homogeneous polynomial of degree $\nu$
(see e.g. \cite{schafer}). There are four infinite series of simple Jordan
algebras, realizable as matrices with the Jordan product being the
anticommutator: the degree 2 algebras $\Sigma(n)$ to be discussed  below, and
the
series $J_k^{\bR}$, $J_k^{\bC}$, $J_k^{\bH}$, which are realized by $k\times k$
hermitian matrices over $\bR$, $\bC$ or $\bH$, with norm given by the
determinant, ${\cal N}(x) =det(x)$. In  addition, there is one
`exceptional' Jordan algebra $J_3^{\bO}$ realizable by $3\times 3$ hermitian
matrices over the octonions.

Associated with any Jordan algebra $J$ with product $x \circ y$ is a
self-dual homogeneous
convex cone ${\cal C}(J)$. This is the subspace of $J$ consisting of elements
$e^x$ with $x\in J$ (where $e^x$ is defined by the usual power series with
$x^{n+1}= x^n \circ x$). The characteristic function is
\be
\omega = {\cal N}^{n/\nu}\, ,
\ee
so the boundary of the cone corresponds to ${\cal N}=0$. The cone is foliated
by
copies of the homogeneous space $Str(J)/Aut(J)$, where $Str(J)$ is the
invariance group of ${\cal N}$ (the `structure group' of the algebra) and
$Aut(J)$ is the automorphism group of the algebra (the subgroup of $Str(J)$
that
fixes the identity element in $J$).

The relation of self-dual homogeneous convex cones to Jordan algebras has
similarities to the relation between Lie groups and Lie algebras. Recall that a
Lie group is parallelizable but has a non-zero torsion given by the structure
constants of its Lie algebra. A self-dual homogeneous convex cone ${\cal C}$,
on the other hand, is not parallelizable (in general) but its torsion-free affine
connection is determined by the structure constants of a Jordan algebra. Because
of homogeneity it suffices to know the connection at the `base' point $c\in{\cal
C}$ defined by\footnote{There is only one such point, even in
those cases for which ${\cal C}$ is flat. It corresponds to the identity
element in the algebra. We use the notation $c$ to indicate both the
identity element of $J$ and the base point of the cone ${\cal C}(J)$.}
\be
g_{ij}|_c=\delta_{ij}\, .
\ee
Let $f_{ij}{}^k$ be the structure constants of $J$ in a basis $e^i=(c,e_a)$.
Then
\be
\Gamma_{ij}{}^k|_c = f_{ij}{}^k\, .
\ee
Although Jordan algebras are commutative they are nonassociative. Define the
{\sl
associator}
\be
\{a,b,c\} \equiv (a\circ b)\circ c - a\circ (b\circ c)\, .
\ee
The curvature tensor of the cone at the base point is then given by the
relation
\be
\{e_i,e_j,e_k\} = R_{ijk}{}^l|_c e_l\, .
\ee

In addition to the automorphism and structure groups, there is a larger
`M{\"o}bius group' associated with any Jordan algebra $J$, acting on
elements of $J$ by fractional linear transformations. We therefore
have the sequence of groups
\be\label{seq}
Aut(J)\subset Str(J) \subset Mo(J)\, ,
\ee
associated with any Jordan algebra $J$. These can be interpreted as
generalized,
rotation, Lorentz and conformal groups, respectively \cite{gunaydin}.
To motivate this interpretation, we return to the representation of a
Minkowski 4-vector as the $2\times 2$ Hermitian matrix $X$.
This is an element in the degree 2 Jordan algebra
$J^{\bC}_2$. The dimension is 4 and the norm is the determinant, which is
the $SL(2;\bC)$ invariant Minkowski norm ${\cal N}$ on $\bR^4$. The group
$SL(2;\bC)$ acts on $2\times 2$ matrices by conjugation so the subgroup leaving
invariant the identity matrix is its maximal compact $SU(2)$ subgroup. The
convex cone associated with this Jordan algebra is the forward light-cone of
D=4 Minkowski spacetime. As we saw in the previous subsection, the
group of fractional linear transformations of $X$ is $SU(2,2)$, so the sequence
(\ref{seq}) is, in this case,
\be\label{seqtwo}
SU(2) \subset SL(2;\bC) \subset SU(2,2)\, .
\ee
These are the standard rotation, Lorentz and conformal groups.

The inclusion of domain wall charges means that we should replace $J^{\bC}_2$
by $J^{\bR}_4$, the algebra of $4\times 4$ symmetric real matrices. One can see
that $J^{\bC}_2$ is a subalgebra of $J^{\bR}_4$ from the fact that $J^{\bC}_2
\cong\Sigma(4)$, where $\Sigma(n)$ is the n-dimensional Jordan algebra with
basis $(1,\sigma_1,\dots \sigma_{n-1})$ and Jordan product $\sigma_a\circ
\sigma_b=2\delta_{ab}$; this has a realization in which $\sigma_a$ are
 sigma-matrices of an $n$-dimensional Minkowski spacetime, with the Jordan
product being the anticommutator; it follows that the standard supersymmetry
algebra in $D$ dimensions is naturally associated with $\Sigma(D)$. For
$D=4$ one can choose the $\sigma_a$ to be the three $2\times 2$ hermitian Pauli
matrices, hence the isomorphism $J^{\bC}_2 \cong\Sigma(4)$. All simple Jordan
algebras of degree 2 are isomorphic to $\Sigma(n)$ for some $n$.
Having replaced $J^{\bC}_2$ by $J=J^{\bR}_4$ we find that the sequence
(\ref{seqtwo}) is generalized to \cite{gunaydin}
\be
SU(2)\times SU(2) \subset SL(4;\bR) \subset Sp(8;\bR)
\ee

We now turn to the Jordan algebraic interpretation of the boundary
of the convex cone ${\cal C}(J)$. This consists of elements $\lambda P\in J$
where $\lambda$ is a positive real number and $P$ is an idempotent of $J$ with
less than maximal rank, i.e. its trace, defined by $\tr X = \log {\cal
N}(e^X)$,
is less than $\nu$. An idempotent is a non-zero element
$P\in J$ satisfying $P\circ P=P$,
and two idempotents $P$ and $P'$ are said to be orthogonal if $P \circ P'=0$.
The idempotents with unit trace are called the primitive idempotents, and the
number of mutually orthogonal primitive idempotents equals the degree $\nu$ of
the algebra. For a Jordan algebra of degree 2 all idempotents of less than
maximal rank have unit trace and are therefore primitive.  This is true of
$J^{\bC}_2$, in particular, corresponding to the fact that the only
supersymmetric states other than the vacuum permitted by the standard D=4 N=1
supersymmetry algebra are 1/2 supersymmetric states associated with massless
particles (for which the 4-momentum lies on the positive light-front). Note
that
although at most two primitive idempotents of a degree 2 Jordan algebra can be
orthogonal in the above sense, the space of primitive idempotents of
$\Sigma(D)$ is
$(D-1)$-dimensional. The boundary of the associated convex cone is therefore
$(D-1)$-dimensional. For $\Sigma(4)\cong J^{\bC}_2$, in particular, this
boundary is the three-dimensional forward light-front of the origin of
4-dimensional Minkowski momentum space.

For a Jordan algebra $J$ of degree $\nu>2$, there are idempotents of less than
maximal rank that are not primitive. For an algebra of degree 3, these
non-primitive idempotents generate faces of the boundary of $e^J$ which
themselves have a boundary generated by the primitive idempotents. An
example is the (non-simple) Jordan algebra $J=\bR\oplus \bR\oplus \bR$ for
which
$e^J$ is the positive octant in $\bE^3$; its boundary consists of three faces
that meet on the three axes generated by the three primitive idempotents (in
this case there are only three primitive idempotents, which are therefore
orthogonal; details can be found in \cite{GST}). More generally, for Jordan
algebras of higher degree, the boundary of the associated convex cone is a
stratified set of faces. In particular, $J^{\bR}_4$ has degree 4 so the faces
of
the boundary of its associated convex cone are generated by idempotents of
trace
1,2 and 3, corresponding to 3/4,1/2 and 1/4 supersymmetry respectively. The
primitive idempotents, of unit trace, correspond to 3/4 supersymmetry.

\subsection{Entropy of BPS fusion} 

In a quantum field theory realization of the D=4 N=1 supersymmetry algebra the
central charges $Z$ are labels of quantum states. We have now seen
that the set of these charges naturally carries the structure of a Jordan
algebra. This algebra may itself be regarded as a finite-dimensional state
space (not to be confused with infinite-dimensional space of states of
the field theory that carry these charges). This interpretation is of course
how Jordan algebras originally arose (see \cite{jordan} for a review). The
exceptional Jordan algebra provides a state space more general than
conventional quantum mechanics but for all other Jordan algebras the formalism
is equivalent to one in which a state is represented by a density matrix. The
general state is therefore a mixed state. The pure states correspond to the
primitive idempotents; these lie on the boundary of the convex cone ${\cal
C}(J)$ but do not in general exhaust it. Rather, the boundary is stratified by
sets of states of successively less purity, corresponding in our application to
states with successively less supersymmetry. Thus, the pure states in this sense
are the charge configurations that preserve 3/4 supersymmetry, the remaining
supersymmetric configurations corresponding to states on the boundary of the
cone that are not pure.

We previously showed that a BPS state is stable against decay into any
other pair of states; in particular it cannot decay into two BPS states.
Consider now the reverse process, i.e. fusion of two BPS states to
form a third via the inverse of the reaction (\ref{fission}), i.e.
\be\label{fusion}
(BPS)_1 + (BPS)_2 \rightarrow (BPS)_3\, ,
\ee
If the first two states preserve 3/4 supersymmetry then the third one will
generally preserve less supersymmetry. This is like passing from a
pure to a mixed state. There is also a formal resemblance here to
classical thermodynamics. The Jordan algebra $J$, now viewed as vector space $V$
containing the convex cone ${\cal C}(J)$, is spanned by the extensive quantities
while the dual vector space $V^*$ is spanned by the intensive variables. The
function
\ben
S(x)= \log \omega (x)
\een
of the extensive variables may be interpreted as entropy. Because it is convex
\ben
S(\mu x +(1-\mu)x') \ge \mu S(x) + (1-\mu) S(x')\, ,
\een
with equality when $x$ is proportional to $x'$, the entropy can not
decrease as a result of a fusion process such as (\ref{fusion}). Conversely, the
(marginal) stability of a single BPS state against decay into two other BPS
states can now be understood as being forbidden by a version of the second law
of thermodynamics.

\section{BPS states for adS}

The N=1 D=4 adS anticommutator (\ref{ads}) may be written as
\be\label{abc}
\{  Q_\alpha, Q_\beta \}={ 1\over 2} M_{AB} \left({\cal C}\Gamma ^{AB}\right)_
{\alpha \beta}, \label{anti}
\ee
where
\be
\Gamma^A=(\gamma^\mu,\gamma_5)\,
\ee
and $M_{AB}=-M_{BA}$ are the generators of the adS group $SO(3,2)$ (and so are
no longer central). The matrix ${\cal C}$ is the $SO(3,2)$ charge conjugation
matrix; we can choose a representation in which  
\be
{\cal C}=\gamma_0\gamma_5
\ee
and this choice will be implicit in what follows. Note that
\be
\{\Gamma^A,\Gamma^B\} = 2 \eta^{AB}\, ,
\ee
where $\eta$ is a flat metric on $\bE^{(2,4)}$, such that
$\eta={\rm diag}(-1,1,1,1,-1)$ in cartesian coordinates. 
Although (\ref{ads}) is preserved by $GL(4;\bR)$, the automorphism group of the
adS supergroup $OSp(1|4;{\bR})$ is $Sp(4;\bR) \subset SL(4;\bR)\subset
GL(4;\bR)$. 

The anticommutator (\ref{ads}) can also be written in the form
(\ref{susyalg}), with
\be
M_{04}=H,\qquad M_{i4}=-P_i,\qquad M_{0i}= -U_i,\qquad J_i \equiv {1 \over 2}
\epsilon_{ijk}M^{jk} =-V_i
\ee
where $H$ is the hamiltonian, ${\bf P}$ the 3-momentum, ${\bf J}$ the angular
momentum while the 3-vector ${\bf U}$ generates boosts.
The analysis of supersymmetric charge configurations is then exactly the same
as in the super-Poincar\' e case considered earlier, and in particular
requiring ${1 \over 4},{1 \over 2}$ or $ {3 \over 4}$ supersymmetry gives
exactly the same conditions on the charges $H,{\bf U,V,P}$ as were found 
earlier. 

The condition for preservation of supersymmetry can be expressed in terms of the
$SO(3,2)$ Casimirs. We will first show how the values of these Casimirs
are constrained by the physical state condition, and then turn to the
supersymmetric states.

\subsection{Physical States in adS}

Physical states lie either in the convex cone for which $Z={1\over 2}  M_{AB}
C\Gamma^{AB}_{\alpha \beta}$ is positive, or on its boundary, for which
${\rm det}Z=0$. This cone is a subspace of the 10-dimensional vector
space spanned by $5\times 5$ skew-symmetric matrices $M$ with entries
$M_{AB}$. The matrix commutator turns this space into the Lie
algebra $so(3,2)$. This algebra has rank 2, with quadratic 
Casimir\footnote{The quadratic Casimir provides a metric of signature $(4,6)$ on
the 10-dimensional vector space, but this metric (which is inherited from
the metric $\eta$ on $\bE^{(3,2)}$) does not play a crucial role in the 
following analysis.}
\be
c_2= {1\over 2}M_{AB} M^{AB}\, ,
\ee
and quartic Casimir
\be
c_4=M^A \,_B M^B\,_C M^C\,_D M^D\,_A .
\ee
Since ${\rm det} Z$ is both a quartic polynomial of the charges and $SO(3,2)$
invariant it must be a linear combination of $c_4$ and $c_2^2$. In fact
\be\label{detz}
{\rm det} Z = c_4-c_2^2\, ,
\ee
and hence
\be
c_4\ge c_2^2\,
\label{fddsgfdg}
\ee
for physical states. 

There is a further constraint on the Casimirs required by physical
states. To see this, we begin by noting that the vacuum is the only physical
state for which the energy $M_{04}$ vanishes. This follows from the
fact that $\{Q,Q\}$ is positive semi-definite, with a trace equal to 
$4M_{04}$. We next prove that $M_{04}$ must vanish if the kernel of $M$ contains
a timelike 5-vector. Suppose that such a 5-vector exists. By an $SO(3,2)$
transformation, we can arrange for it to have only one non-vanishing component,
in the $4$-direction. It then follows that the only non-vanishing components of
$M$ are $M_{\mu\nu}$. In particular, the energy $M_{04}$ vanishes. Thus, {\sl
for any non-vacuum physical state the kernel of $M$ contains no timelike
vectors}. Note that the kernel of $M$ has dimension $1$, $3$ or $5$, according
to whether $M$ has rank $4$, $2$ or $0$, respectively. The vacuum is
the only physical state for which $M$ has rank $0$. 

Now consider the Pauli-Lubanski 5-vector
\be
s^A= { 1\over 8} \epsilon^{ABCDE}M_{BC} M_{DE}\, .
\ee
This satisfies the identity
\be\label{idens}
M_{AB}s^B\equiv 0\, ,
\ee
which shows that, unless it vanishes, $s$ is in the kernel of $M$. A timelike
$s$ would therefore be in the kernel of $M$ but, as we have just seen,
the kernel of $M$ cannot contain timelike vectors unless $M$ vanishes,
but in that case $s$ also vanishes. Thus, $s$ cannot be timelike. Now,
\be\label{vsquared}
s^2\equiv \eta^{AB}s_As_B = {1\over 4}(2 c_2^2-c_4)\, ,
\ee
so $s$ will be non-timelike if and only if  
\be\label{upperb}
c_4 \le 2c_2^2\, .
\ee
This bound implies (for physical states) that $c_4=0$ when $c_2=0$ .

\subsection{Supersymmetric States}

Our main interest is in BPS states, i.e. the subset of physical states that
are supersymmetric. These must saturate the bound (\ref{fddsgfdg}), so BPS
states are those for which
\be
c_4=c_2^2\, .
\ee
Using this in (\ref{vsquared}) we see that
\be
s^2 = {1\over4}c_2^2
\ee
for {\sl supersymmetric} states. We will organise our discussion of the
supersymmetric states according to whether $s$ is zero, spacelike
or non-vanishing null. 

If $s$ vanishes then $M$ has either a 3-dimensional or a 5-dimensional kernel.
$M$ will have a 5-dimensional kernel only if it vanishes. If the kernel is
3-dimensional then, as we have seen, it cannot contain timelike vectors. It may
contain null vectors but any such null vector must be orthogonal to all other
vectors in the kernel, spacelike or null, because we could otherwise find a
timelike linear combination. Since the maximum number of mutually orthogonal null
5-vectors is $2$, a 3-dimensional kernel must contain at least one spacelike
vector. There are three possible choices for the other two linearly independent
5-vectors:  (i) both spacelike, (ii) one spacelike and one null, or (iii) both
null. In all cases $M$ can be brought to a form in which $M_{04}=E\ge0$ is its
only {\sl independent} entry. In case (i) $M_{04}$ and
$M_{40}$ are the only entries, and the only {\sl supersymmetric} state with this
property is the vacuum, with $E=0$. In case (ii) $M$ can be brought to a form
for which the only non-zero upper-triangular entries are $M_{04}=M_{02}=E$. It
then follows from the discussion of section 2.4, on which we will elaborate
below, that {\sl all} such states are 1/2 supersymmetric. In case (iii) $M$ can
brought to a form for which the only non-zero upper-triangular entries are
$M_{04}=-M_{02}=M_{23}=M_{34}$; {\sl all} such states are 3/4 supersymmetric. 

Consider now spacelike $s$. In this case we may choose the only non-vanishing
component of $s$ to be its $1$-component. Since $s$ now spans the kernel of
$M$, this $5\times 5$ matrix $M$ then reduces to a 
$4\times 4$ matrix $F$ acting
on the 4-dimensional $(0234)$ subspace orthogonal to $s$, on which
$\eta$ restricts to a metric $\tilde \eta$ of signature $(2,2)$. The matrix $F$
is equivalent to a second-rank antisymmetric tensor in $\bE^{(2,2)}$ 
that can be
written uniquely as $F=F^+ + F^-$ where $F^+$ is 
real and self-dual while $F^-$
is real and anti-self-dual matrix.  Now
\be
c_4 -c_2^2 = \left[{\rm tr} (\tilde\eta F^+)^2\right]\left[{\rm tr} (\tilde
\eta F^-)^2\right]\, .
\ee 
We can write $F$ as
\be\label{eff}
F=\pmatrix{0&u&b&E\cr -u&0&-v&c\cr -b&v&0&-p\cr -E&-c&p&0}
\ee
provided that 
\be
vE + bc +up \ne 0\, ,
\ee
since $s$ would otherwise vanish. Now
\be
-{\rm tr} (\tilde\eta F^\pm)^2 = (E\mp v)^2 - (u\pm p)^2 -
(b\pm c)^2\, .
\ee
Configurations with self-dual or anti-self-dual $F$, for 
which $E=\mp v$, $u=\pm p$ and $b=\pm c$, are 1/2 supersymmetric.  
However, any configuration for which
\be\label{qwerty}
(E \mp v)^2 = (u \pm p)^2 + (b\pm c)^2
\ee
is also supersymmetric. In fact
\be\label{tubby}
\{Q,Q\} = \left[(E\mp v) - (b\pm c)\gamma^{012} + (u\pm p) \gamma^{013}\right]
+ \left( v-c\gamma^{02} + p\gamma^{03}\right)\left(1\pm \gamma^1\right)\, .
\ee
Given (\ref{qwerty}), the term in square brackets is proportional to a 1/2
supersymmetry projector that commutes with the 1/2 supersymmetry projector
$(1/2)\left(1\pm \gamma^1\right)$ which leads generically to 1/4 supersymmetry.

The final case to consider is $s$ null but non-zero. By means of an $SO(3,2)$
transformation we may choose 
\be\label{ess}
s\propto (1,0,1,0,0)
\ee
This choice is preserved by an $SO(1,2)$ `stability' subgroup, and by a
transformation in the $SO(2)$ subgroup of this group we can bring $M$
to the standard form   
\be\label{Mform}
M=E\pmatrix{0&0&-a&0&1\cr 0&0&a&0&-1\cr a&-a&0&t&-q\cr 
0&0&-t&0&-r\cr -1&1&q&r&0}
\ee
One then finds that
\be
c_2 = E^2(t^2-q^2 -r^2)\, ,
\ee
so that supersymmetric states are those with 
\be\label{susy1}
t= \pm\sqrt{q^2+ r^2}\, .
\ee
Actually, in arriving at the above form of $M$ we have used only that the null
5-vector $(1,1,0,0,0)$ is in the kernel of $M$. To ensure that this 5-vector 
is proportional to $s$ (with non-zero constant of proportionality) we require
that
\be\label{susy2}
t+ra\ne 0\, .
\ee
This condition also ensures (as it must) that $M$ has rank 4. When combined
with (\ref{susy1}) it implies that
\be\label{susy3}
t\ne0\, .
\ee

For $M$ of the form (\ref{Mform}) we have
\be\label{QQ}
\{Q,Q\} = E\left\{ \left(1-a\gamma_3\right)\left(1-\gamma_{01}\right) 
-t\gamma_1\left[1 - (q/t)\gamma_{012} - (r/t) \gamma_{013}\right] \right\}
\ee
A spinor $\epsilon$ is in the kernel of $\{Q,Q\}$ if
\be
\left[(q/t)\gamma_{012} + (r/t)\gamma_{013}\right]\epsilon =\epsilon\, ,
\ee
and
\be
\gamma_{01}\epsilon=\epsilon\, ,
\ee
and these two constraints imply 1/4 supersymmetry. Note that 
when $a=\pm 1$ and $q=0$ and hence $t=\pm r$,
the latter constraint can be replaced by $\gamma_3\epsilon=\pm 
\epsilon$, which again yields 1/4 supersymmetry. 

\subsection{Examples}

Many of the possibilities for BPS configurations just noted are illustrated by
the class of examples considered in section 2.4. This means, in the language of
this section, that the non-zero upper-triangular components of $M_{AB}$ are 
taken to be $M_{04}=E$, $M_{34}=-p$, $M_{02}=u$ and $M_{23}=-v$. 
The Pauli-Lubanski 5-vector is then 
\be
s=(0,Ev+up,0,0,0)\, ,
\ee
so $s$ is spacelike unless it vanishes. The Casimirs for this class 
are given by
\bea
c_2&=&E^2+v^2-p^2-u^2\\
c_4&=&2[E^4+u^4+v^4+p^4-2(v^2+E^2)(u^2+p^2)-4Euvp]
\eea
The BPS condition $c_4=c_2^2$ becomes
\be\label{BRStest}
(E-u-v-p)(E-u-v+p)(E-u+v-p)(E+u-v-p) =0\, ,
\ee
in agreement with (\ref{eigen}). 

Let us first consider vanishing $s$. We have seen above that $M$ can be brought
to a standard form in which all charges are determined in terms of $M_{04}=E$.
The non-vacuum BPS states occured for cases (ii) and (iii) discussed above.
An example of case (ii) within the class of configurations now under discussion
is found by setting $v=p=0$ and $E=|u|$\footnote{The charge $u$ 
can be interpreted as a membrane charge. To see this note that
there is a static planar solution of the equations of motion 
of a test membrane in $adS_4$ at a fixed radial distance, in horospherical 
coordinates, from the Killing horizon \cite{kal}. This solution must
preserve 1/2 supersymmetry of the $adS_4$ supersymmetry because $adS_4$
can itself be interpreted as a membrane, at the horizon, to which the 
test membrane is parallel. Because this test membrane remains at a fixed 
distance from the horizon, the worldline of a point on it is uniformly 
accelerated, and therefore naturally associated with a non-zero boost $u$.}. 
Finally, an example of case (iii), with 3/4 supersymmetry, is obtained by  
setting $u=v=p=-E <0$, although there is no known field theoretic 
realization of this case. 

We next to turn to examples with $s$ spacelike. Let us first 
consider $u=p=0$ and set $v=-J$, where $J$ is the spin about the 1-axis.
We then have
\be\label{c2c4}
c_2=E^2 + J^2\, ,\qquad c_4=2E^4 + 2J^4\, ,
\label{Paulis}
\ee
which is equivalent to
\bea
E &=& \sqrt{ { c_2 \over 2} + {1\over 2}\sqrt{ c_4- c_2^2}}\, ,\nonumber\\
J &=& \pm \sqrt{ { c_2 \over 2} - {1\over 2} \sqrt{ c_4- c_2^2 }}\, .
\eea
The physical states satisfy $E\ge|J|$ and states that saturate this bound
preserve 1/2 supersymmetry. For these configurations the matrix $F$ of
(\ref{eff}) is either self-dual or anti-self-dual. 
An example of states with
$s$ spacelike and $F$ neither self-dual or anti-self-dual
can be obtained by taking $u,v,p$ to be positive and solving
(\ref{qwerty}) via $E=u+v+p$. We then have
\be\{Q,Q\} = u(1+\gamma^{013})+p(1+ \gamma^{03}) +v(1+\gamma^1)
\ee
and 1/4 of the supersymmetry is preserved.

\section{Comments}

We have seen that a full analysis of the D=4 N=1 supersymmetry algebra not only
confirms the existence of 1/2 and 1/4 supersymmetric states, realizable within
the WZ model, and determines some of their properties, but it also permits
states with 3/4 supersymmetry \cite{GH,Bandos,Ueno} which, as we have shown,
cannot be realized by solutions of 
the WZ model. However, it has been argued that such
`exotic' fractions might play a role in other contexts, and with this in mind
we have provided a detailed analysis of the BPS states of 
D=4 N=1 supersymmetry.
We have also seen that these states can be understood in terms of the geometry
associated with the convex cone of the Jordan algebra $J_4^{\bR}$, and that
this leads to a natural generalization of the rotation, Lorentz and conformal
groups.

In general, the $U(1)_R$ 
symmetry will be broken  to at most a
discrete subgroup. 
For 
theories with domain walls (e.g. the WZ model), the R-symmetry will  be
explicitly broken by the scalar potential. In theories with only massless
particles, and  no domain walls,  the $U(1)_R$ symmetry will be generically
broken to a discrete subgroup  by chiral anomalies.
For theories in which the domain wall charges are quantized, the 
$U(1)_R$ symmetry will be broken to the discrete subgroup preserving the
quantization condition. 
 An example of this is given by
M-theory compactified on a 7-manifold  of $G_2$ holonomy,
yielding a D=4 N=1 theory in which the domain walls are M2-branes and 
wrapped M5-branes, with the M2-brane and M5-brane charges   quantized.
Given that only a discrete subgroup of $U(1)_R$ survives the same is 
true of the larger group $SU(2)_R$. 

We noted that, in the classical theory, the automorphism group of the full
supertranslation algebra is  $GL(4,\bR)$, but it seems that any realization of
this on fields, and any realisation of the generalized conformal group
$Sp(8,\bR)$, requires an enlargement of 3-space to include coordinates
conjugate to the `domain-wall' charges ${\bf U}$ and ${\bf V}$. Of course, the
domain wall interpretation is probably no longer appropriate in this case.
Other interpretations are certainly possible in the context of particle
mechanics \cite{BLS}. In such one-dimensional field theories it is possible to
realize the $SU(2)_R$ symmetry between the three 3-vector `charges' ${\bf P},
{\bf U}, {\bf V}$  as an internal symmetry.
 For such models that
arise  from the toroidal compactification of some D=4 theory with quantized
${\bf U}$ and ${\bf V}$, the 3-momentum will also be quantized and the
classical  $GL(4;\bR)$ symmetry will be broken to the discrete $GL(4;\bZ)$
subgroup preserving the 9-dimensional charge lattice.

Many of the observations made here for $N=1$ $D=4$ can of course be generalized
to $N>1$ or to $D>4$. For example the general $N$ extended $D=4$ supersymmetry
algebra has automorphism group $GL(4N;\bR)$ and $\det\{ Q,Q \}$ is preserved by
the subgroup $SL(4N, \bR )$. This leads to the
sequence
\be\label{seqetc}
SO(4N) \subset SL(4N;\bR) \subset Sp(8N;\bR)\, .
\ee
for the Jordan algebra $J_{4N}^{\bR}$ of $4N \times 4N $ symmetric matrices
over the reals. The generalised conformal symmetry of the BPS condition is then
$Sp(8N;\bR)$, as deduced from a different analysis in \cite{FP}. 

A $D>4$ case of particular interest is the D=11 `M-theory algebra' $\{Q,Q\}=Z$
where $Q$ is now a 32 component real spinor of the D=11 Lorentz group and $Z$ is
a $32\times 32$ real symmetric matrix containing the Hamiltonian and 527 central
charges carried by M-branes \cite{PKT}. This supersymmetry algebra has
automorphism group $GL(32;\bR)$, as noted independently in \cite{bw}, and $Z$
takes values in the convex cone associated with the Jordan algebra
$J_{32}^{\bR}$. The sequence (\ref{seq}) of groups associated with this
algebra is 
\be
SO(32) \subset SL(32;\bR) \subset Sp(64;\bR)\, ,
\ee
so that $Sp(64;\bR)$ is the M-theoretic generalisation of the D=11 conformal
group. 
As in the D=4 case, the realization of any of these larger `spacetime' symmetry
groups, or discrete subgroups such as $GL(32;\bZ)$, would seem to require
consideration of an enlarged space of 527 coordinates, as considered for other
reasons in \cite{Hull}. 

Finally, we have found many possibilities for new BPS states in anti de Sitter
space. It seems likely that some of these, in particular those with 1/4
supersymmetry, will have a realization in the context of N=1 D=4 
supersymmetric field theories in an adS spacetime.

\medskip
\section*{Acknowledgments}
\noindent
We would like to thank C. Gui for bringing ref. \cite{gui} to our attention.
We also thank M. G\"unaydin and J. Lukierski for helpful correspondence. JPG
thanks the EPSRC for partial support.
The work of CMH was supported in part by the National Science
{}Foundation under Grant No. PHY94-07194. All authors are supported
in part by   PPARC through their SPG $\#$613.


\end{document}